\documentclass[aps,eqsecnum,groupedaddress,a4paper,floatfix,twocolumn]{revtex4}
\usepackage[dvips]{graphicx}
\usepackage{latexsym}

\begin{document} 

\title{Limits to phase information in the parametric oscillator}

\author{P. Kinsler} 
\altaffiliation[This research was primarily done at the ]{
Department of Physics, University of Queensland,
Australia 4072, as part of my PhD in 1993-1994, and was largely
contained within my thesis\cite{Kinsler-1994}}
\email{Dr.Paul.Kinsler@physics.org}

\affiliation{
Department of Physics, Imperial College,
  Prince Consort Road,
  London SW7 2BZ, 
  United Kingdom}

\begin{abstract} 

Light in which the quantum fluctuations have been squeezed is often proposed as
a means of obtaining an improved phase reference compared to that available
from coherent light.  Such a phase reference contains information about the
phase of the squeezed light, so it is important to calculate the limits to the
amount of {\lq\lq}phase information{\rq\rq} available.  I define a phase
resolution and show how this scales as we increase the number of photons
available when using squeezed light generated by a parametric oscillator. 
Simple schemes for creating squeezed coherent light using a beam-splitter and
an interferometer are analyzed, and it is shown that the results agree with
earlier claims\cite{Kinsler-FD-1993,Kinsler-D-1995,Kinsler-1994} that when
using squeezed coherent light as a phase reference, the gain is not one of
improved accuracy, but of lower power

\end{abstract}

\date{\today} \maketitle




\begin{section}{Introduction}

The amount of squeezing that can be generated is of interest in many potential
applications, ranging from high resolution
spectroscopy\cite{Heidmann-HRGFC-1987} to gravity wave
detection\cite{Hollenhorst-1979,Caves-1981}.  In these, the reduced quantum
fluctuations (squeezing) in one quadrature of a light field are used to obtain
a better signal to noise ratio than that given by the standard quantum limit. 
In this paper I consider a commonly suggested scheme for generating squeezed
light, the use of a parametric amplifier, where a high frequency
{\lq\lq}pump{\rq\rq} field is down converted into half-frequency sub-harmonic
fields (often called the {\lq\lq}signal{\rq\rq} and {\lq\lq}idler{\rq\rq}).  In
calculations for an idealised lossless parametric
amplifier\cite{Kinsler-FD-1993,Hillery-Z-1984,Scharf-W-1984,Crouch-B-1988}, and
for the more realistic non-equilibrium calculations at the threshold of
parametric oscillation\cite{Kinsler-D-1995}, the squeezed quadrature variance
in the sub-harmonic field(s) is found to scale as $N^{-1/2}$, where $N$ is the
number of photons in the pump field. Crouch and Braunstein\cite{Crouch-B-1988}
suggested on intuitive grounds that  for the lossless parametric oscillator,
the {\lq\lq}phase uncertainty{\rq\rq} in the pump field is a lower bound on the
phase uncertainty in the sub-harmonic field. Since the phase variance in the
initial coherent state in the pump is of the order of $N^{-1}$, they inferred
an approximate lower bound for the variance of the squeezed quadrature which
varies as $N^{-1/2}$.  

This result of $N^{-1/2}$ scaling in the variance of the squeezed
quadrature implies a $N^{1/2}$ scaling in the unsqueezed fluctuations. 
Consequently, the efficiency of energy transfer from the N-photon pump
field to the signal field scales with $N^{-1/2}$, since the unsqueezed
fluctuations contain most of the energy. Kinsler, Fernee and
Drummond\cite{Kinsler-FD-1993} conjectured that this bound on efficiency is
universal when a coherent field and a phase-invariant Hamiltonian combine
to form a squeezed vacuum. The reason for this is straightforward. We
should not be able to infer more information about the pump phase than its
inherent variance of $N^{-1}$, its standard deviation divided by its
amplitude.  This is precisely the phase uncertainty of an ideal squeezed
vacuum that has been generated with a relative efficiency of $N^{-1/2}$.  

In the applications mentioned above, there are often different
strategies which involve utilising increased coherent power levels
rather than squeezed light to obtain greater precision
\cite{Caves-1981}. If the production of squeezed light itself requires
large coherent power inputs, then it may be advantageous to use a
strategy involving only an increased coherent power input -- assuming
the equipment can tolerate the higher powers. As a result, the power
input required to produce a known degree of squeezing is a significant
factor.  

\end{section}

\begin{section}{Phase resolution}

Squeezed light is often suggested as a way to obtain an improved phase
reference. To see if this is practical, we need to consider the amount of
squeezing produced and the efficiency of its production, as well as
defining some measure of how good this phase reference is.  To this end I
define a phenomenological measure of the phase information available from
an optical state: the phase resolution $S$. For example, with a coherent
state, for a larger field amplitude $\alpha$ we get a better defined phase
-- this is because the uncertainty in the position of the amplitude is
always the same (the standard quantum limit of 1), but $\left| \alpha
\right|$, the distance from the origin, has increased.  For $\left| \alpha
\right| \gg 1 $, the phase resolution is 

\begin{eqnarray}
S_{coh} = \frac
            {{\rm distance\ from\ origin}}
            {{\rm uncertainty\ in\ position}}
        =  \frac{\left| \alpha \right|}{1} 
        = N^{1/2}, 
\label{def-phase-resolution-coh}
\end{eqnarray}

since there are $N$ photons in a coherent state with amplitude $\alpha$.  This
is the reciprocal of the phase variance.  Note that Freyberger and Schleich
have considered a variety of definitions of phase in a more comprehensive
fashion\cite{Freyberger-S-1994}. A quantum mechanical approach such as that
introduced by Pegg and Barnett\cite{Pegg-B-1989} could also be used, but
quantum phase effects are unlikely to be important in the large photon number
limit considered here.

For more general states, such as squeezed states, we need a more general
definition.  We can get this by defining two quadratures of the optical
state.  The first quadrature $\hat{Y}$ is the {\lq\lq}distance{\rq\rq}
quadrature, oriented along the line between the origin and the expectation
value of the amplitude.  For a coherent state, $\left| \left< \hat{Y}
\right> \right| = \left< \hat{Y}^\dagger \hat{Y} \right>^{1/2} = \left|
\alpha \right| = N^{1/2}$.  The second is the {\lq\lq}uncertainty{\rq\rq}
quadrature $\hat{X}$, which is oriented orthogonal to $\hat{Y}$.  The
variance of $\hat{X}$ is the usual $\left< \Delta \hat{X}^2 \right> =
\left< \hat{X}^2 \right> - \left< \hat{X} \right>^2$.  Using this approach,
we can define a phase resolution of

\begin{eqnarray}
S = \left[ \frac {   \left<\hat{Y}^\dagger \hat{Y}\right> }
                 {   \left<   \Delta \hat{X}^2\right    > }
    \right] ^{1/2}
,
\label{def-phase-resolution}
\end{eqnarray}

which is large for a well-defined phase. The situtaion is shown pictorially in
Fig. \ref{Fig1}, which corresponds to a phase space picture derived from the
Wigner representation\cite{Wigner-1932}.  This definition also works for
squeezed vacuums, which have a zero average field but a non-zero intensity that
increases with the degree of squeezing.  Consequently, the greater the
squeezing, the better defined (to within $\pi$) is the orientation of the
squeezed state.

\begin{figure}
\includegraphics[width=80mm]{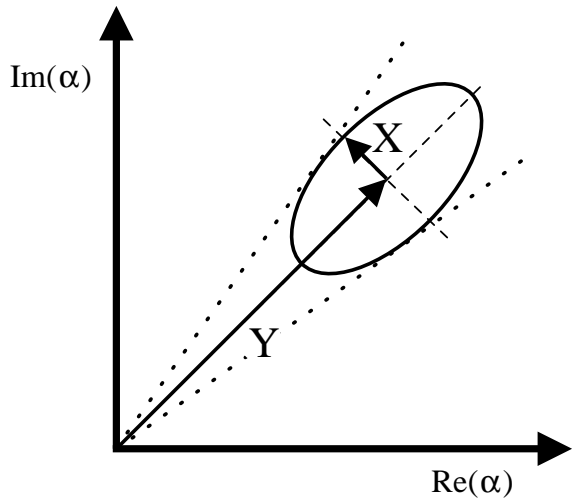}
\caption{ 
\label{Fig1}
The phase uncertainty of a squeezed coherent state.  The ellipse 
represents the contour on which the Wigner distribution function of the 
state has {\em half} its peak value.  The distance quadrature $\hat{Y}$, 
and uncertainty quadrature $\hat{X}$ are represented by arrows. 
The ratio of $\hat{Y}$ to $\hat{X}$ gives an estimate of how well
the phase of the state is defined, as indicated by the angular 
separation of the dotted lines. 
}
\end{figure}

\end{section}

\begin{section}{The parametric oscillator}

Optical parametric oscillators are one of the most interesting and well
characterized devices in nonlinear quantum optics.  Novel discoveries made with
them include demonstrations of large amounts of squeezing\cite{Wu-KHW-1986},
significant quantum intensity correlations\cite{Heidmann-HRGFC-1987}, and a
quadrature correlation measurement that provided the first experimental
demonstration of the original EPR paradox\cite{Ou-PK-1992}.  In addition, other
work\cite{Kinsler-D-1995,Drummond-K-1995} has presented solutions for the
at-threshold behaviour of the degenerate and non degenerate systems, when they
are far from equilibrium and dominated by quantum mechanical effects. 
Practical applications include their use as highly efficient and tunable
frequency converters\cite{Spec}.  

The degenerate parametric oscillator is an idealised interferometer,
which is resonant at two frequencies, the sub-harmonic (or signal)
frequency $\omega_1$, and the pump field frequency $\omega_2 =
2\omega_1$.  Both fields are damped due to cavity losses.  It is
externally driven by a laser field tuned to the pump frequency.  Pump
photons are down converted to pairs of resonant sub-harmonic photons due
to a $\chi^{(2)}$ nonlinearity present inside the cavity; and the
reverse process where a pair of sub-harmonic photons combine to form a
pump photon also occurs.  The non degenerate parametric oscillator is
similar, except that there are two lower frequency fields, called the
signal and idler. Pump photon are down converted into a signal--idler
pair of photons, and as before, the reverse process also occurs.

From eqn (\ref{def-phase-resolution}), the phase resolution for the
sub-harmonic (or signal and idler) fields for the degenerate ($i=2$) and non
degenerate ($i=3$) parametric oscillators are denoted $S_2$ and $S_3$
respectively, are

\begin{eqnarray}
S_i &=& \left[ { \left< \hat{Y}_i^\dagger \hat{Y}_i \right>
                 \over
                 \left< \Delta X_i^2 \right>
               }
        \right] ^{1/2}.
\label{def-signal-to-noise}
\end{eqnarray}

If $\hat{a}$ is the annihilation operator for the sub-harmonic field mode
of the degenerate parametric oscillator, the {\lq\lq}distance{\rq\rq}
quadrature is $\hat{Y}_2 = \hat{a}_1 + \hat{a}_1^\dagger$.  The
{\lq\lq}uncertainty{\rq\rq} quadrature $X_2$ is defined as $\hat{X}_2 = -i
(\hat{a}_1^\dagger - \hat{a}_1)$\cite{Drummond-MW-1981}.  Similarly, in the
non degenerate system, if $\hat{a}_2$ and $\hat{a}_3$ are annihilation
operators for the signal and idler modes, the  {\lq\lq}distance{\rq\rq}
quadrature is a combination of $\hat{a}_2$ and $\hat{a}_3$, being $Y_3 =
\hat{a}_+ = i(\hat{a}_2 - \hat{a}_3)$.  The uncertainty quadrature $X_3$ is
defined as $\hat{X}_3 = -\imath (\hat{a}_+^\dagger - \hat{a}_+)$.  These
definitions are equivalent to those used in\cite{Kinsler-FD-1993}.


\begin{figure}
\includegraphics[width=80mm]{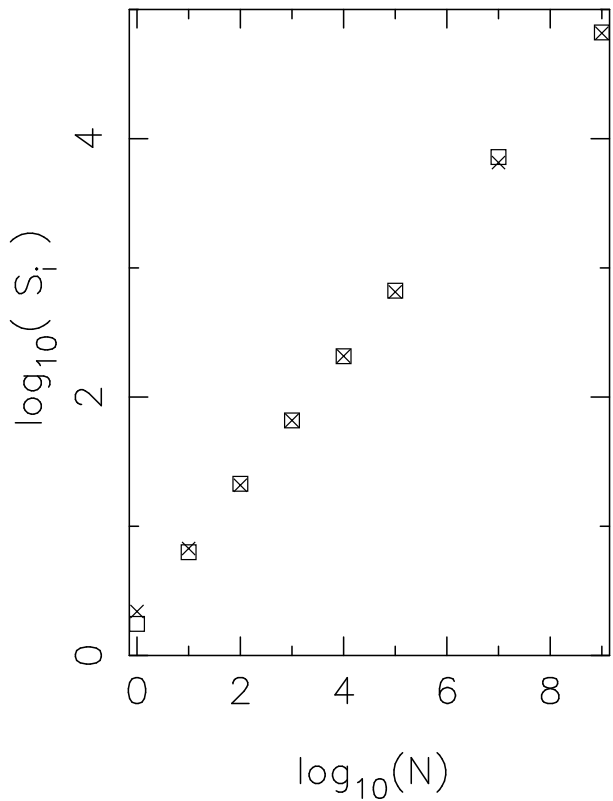}
\caption{
\label{Fig2}
Phase resolution of the squeezed state inside a lossless parametric 
oscillator. The data is from the simulations presented in 
\protect\cite{Kinsler-FD-1993,Kinsler-1994}. 
The graphs show the phase resolutions for the squeezed vacuum state 
produced by 
a degenerate ($\times$) and non degenerate ($\Box$) system.  
}
\end{figure}

Consider a parametric oscillator with perfect mirrors.  At $t=0$ it has a
coherent state in the pump mode, and vacuum(s) in the sub-harmonic mode(s).  As
the system evolves in time, the energy oscillates between the modes; and at a
certain time $t_{sq}$, the squeezing in the sub-harmonic mode(s) is
maximised\cite{Kinsler-FD-1993}.  Figure \ref{Fig2} shows the numerically
calculated optimum values of the phase resolution for this case as a function
of the number of photons $N$ initially in the pump mode.  The phase resolution
scales as $N^{1/2}$ for both types of parametric oscillator.  This is just as
expected because the (squeezed) uncertainty $\left< \Delta \hat{X}_i^2 \right>$
scales as $N^{-1/2}$, and the (intensity) distance $\left< \hat{Y}_i^\dagger
\hat{Y}_i \right>$ scales as $N^{1/2}$ (and $\left[ N^{1/2} / N^{-1/2}
\right]^{1/2}=N^{1/2}$).  Clearly this phase resolution scales with $N$ in
exactly same fashion as it would for the coherent state used as the initial
condition.

Defining a phase resolution for the non equilibrium case of a driven (and
lossy) parametric oscillator involves a little more thought, as we need to
measure the output fields, not the internal cavity fields. The output squeezing
spectrum $\tilde{V}_{i-out}(\omega)$ is the non equilibrium counterpart to the
uncertainty quadrature $\hat{X}_i$.  So what is the counterpart to the distance
quadrature $\hat{Y}_i$? The two candidates are either the steady state output
intensity $\bar{W}_{i-out}$, and the intensity spectrum
$\tilde{W}_{i-out}(\omega)$.  The latter choice gives a consistent definition,
that of the phase resolution $\tilde{S}_{i-out}(\omega)$ of a particular
frequency component.  For the degenerate ($i=2$) and non degenerate ($i=3$)
cases this is

\begin{eqnarray}
\tilde{S}_{i-out}(\omega) &=& \left[ 
                 {    \tilde{W}_{i-out}(\omega)
                 \over
                      \tilde{V}_{i-out}(\omega) 
                 }
        \right] ^{1/2},
\label{def-output-signal-noise}
\end{eqnarray}


where the spectra of the (squeezed) uncertainty
and (unsqueezed) distance quadratures are\cite{Gardiner-1991QN}

\begin{eqnarray}
\tilde{V}_{i-out}(\omega) &=& {4 \pi \gamma \over T} 
  \left< : \tilde{X_i}^\dagger(\omega), \tilde{X_i}(\omega) : \right> + 1
,
\nonumber
\\
\tilde{W}_{i-out}(\omega) &=& {4 \pi \gamma \over T} 
  \left< : \tilde{Y_i}^\dagger(\omega) \tilde{Y_i}(\omega) : \right> + 1
.
\label{def-output-spectra}
\end{eqnarray}

The tilde is used to denote the Fourier transformed quadratures, and the
colons(:) denote normal ordering. Figure \ref{Fig3} shows the numerical
calculations of $\tilde{S}_{i-out}(\omega)$, worked out near threshold since
this is where the best squeezing is obtained.  These scale as $N^{1/2}$,
exactly as would the same quantity calculated for a simple coherently driven
cavity containing $N$ photons.  

\begin{figure}
\includegraphics[width=80mm]{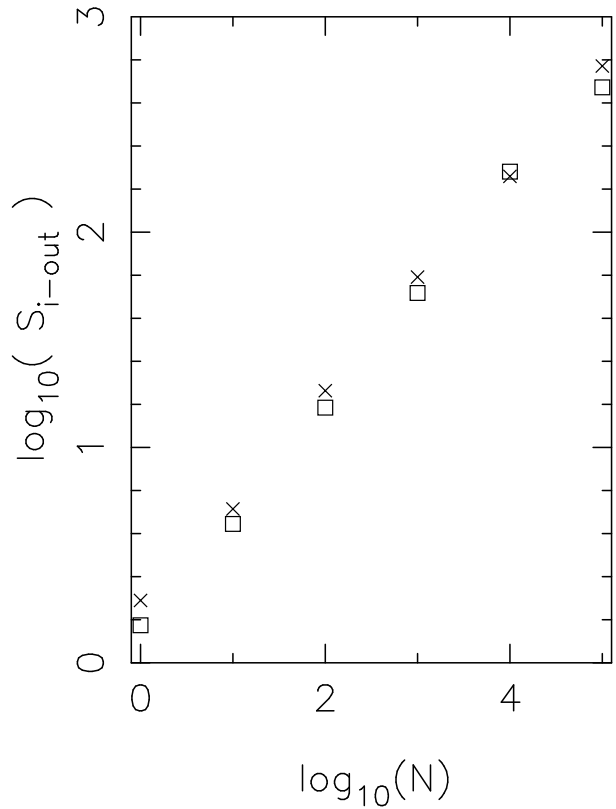}
\caption{
\label{Fig3}
Phase resolution of the output from a driven parametric oscillator. 
The data is from the simulations presented in 
\cite{Kinsler-D-1995,Kinsler-1994}
and are at threshold in 
order to get the best possible squeezing. The graphs show 
the phase resolutions for the squeezed vacuum 
state produced by 
a degenerate ($\times$) and non degenerate ($\Box$) system.  
}
\end{figure}

For a squeezed vacuum, the intensity spectrum $\tilde{W}_{i-out}(\omega)$ in
eqn (\ref{def-output-signal-noise}) could be replaced by the spectral variance
in the unsqueezed quadrature. This would give a definition entirely in terms of
the quadratures, which might seem better than the one used above.  However, the
zero frequency part of the unsqueezed spectra $\tilde{W}_{i-out}(0)$ scales not
as $N^{1/2}$, but as $N$\cite{Kinsler-D-1995}. This quadrature-only definition
would lead to a phase resolution that scaled as $N^{3/4}$ ($ = \left[ N /
N^{-1/2} \right]^{1/2}$).  At first sight this has exceeded that of the coherent
driving field which is the input to the system.  This apparent gain is due to
the slowing down in the critical fluctuations that occur in the unsqueezed
quadrature near threshold. These fluctuations vary over a time that is
$N^{1/2}$ longer than those in the squeezed quadrature, and as a result a phase
resolution based on this quadrature-only definition has a time averaging built
into it. The gain occurs at the expense of ignoring what is going on in the
other frequency components of the output light.  When an integration over all
frequency components is added, the $N^{1/2}$ scaling would be recovered.

\end{section}
\begin{section}{Making squeezed coherent light}

It is possible to combine the squeezed vacuum with a coherent field to
produce a squeezed field with a coherent amplitude by using a beam-splitter
or interferometer. Here I consider whether this can improve the phase
resolution given the $N^{-1/2}$ squeezing efficiency limitation.  

\begin{figure}
\includegraphics[width=80mm]{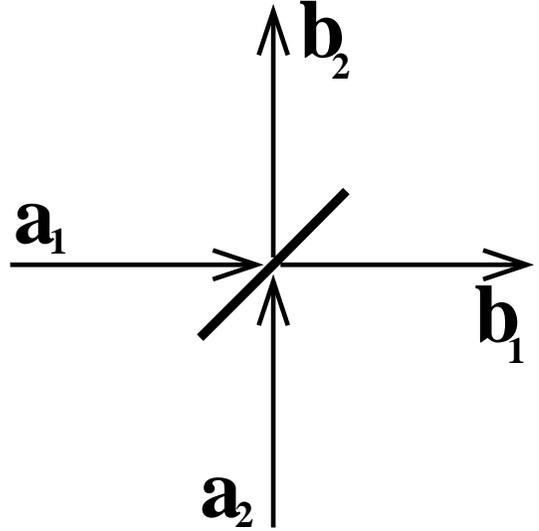}
\includegraphics[width=80mm]{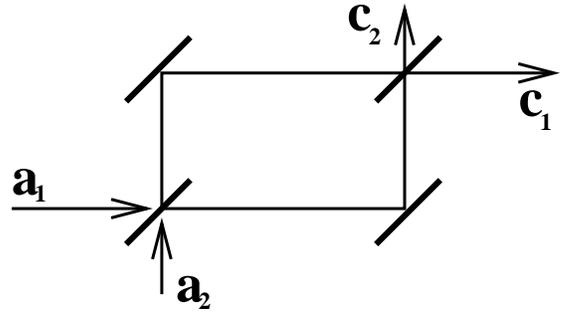}
\caption{
\label{Fig4}
Diagrams of (a) a beam-splitter, and (b) an interferometer, 
showing the input and output modes.
}
\end{figure}

A schematic diagram of a beam-splitter is given in Fig. \ref{Fig4}(a). The
two input modes are $\hat{a}_1$ and $\hat{a}_2$, and these are coupled by
the beam-splitter to the output modes $\hat{b}_1$ and $\hat{b}_2$. The
coefficients of transmission and reflection of the beam-splitter are given
by $t_i$, $r_i$. The equations coupling the two sets of modes are 

\begin{eqnarray}
\hat{b}_1 &=& { \exp}(i\Delta) \left[ t_1 \hat{a}_1 
            + { \exp}(i \psi) r_2 \hat{a}_2 \right]
\nonumber \\
\hat{b}_2 &=& { \exp}(i\Delta) \left[ t_2 \hat{a}_2 
            - { \exp}(i \psi) r_1 \hat{a}_1 \right]. 
\label{def-bs-coupling}
\end{eqnarray}

Here $\Delta$ is an overall phase shift, and $\psi$ is a relative phase
shift. The input states to this beam-splitter will be a coherent state
$\left| \alpha \right>$ in $\hat{a}_1$, and a squeezed vacuum $\left| 0, s
e^{i \theta} \right>$ in $\hat{a}_2$, and we want to analyze the squeezed
coherent light in the $\hat{b}_1$ output mode.  The uncertainty quadrature
$\hat{X}_{BS} = \hat{b}_1 +  \hat{b}_1^\dagger$ has a variance given by

\begin{eqnarray}
\left< \Delta \hat{X}_{BS}^2 \right> = 
1 + 2 r_2^2 \sinh (s) 
\left[ 
    \sinh (s) - \cosh (s) \cos( 2 \Delta + 2 \psi + \theta ) 
\right]. 
\label{def-bs-variance}
\end{eqnarray}

The optimum amount of squeezing occurs for $2\Delta+2\psi+\theta=0$, and is

\begin{eqnarray}
\left< \Delta \hat{X}_{BS}^2 \right> = 
1 - r_2^2 \left[ 1 - { \exp}(-2s) \right].
\label{eqn-bs-squeezing}
\end{eqnarray}

The distance quadrature $\hat{Y}_{BS} = \hat{b}_1$, so

\begin{eqnarray}
\left< \hat{Y}_{BS}^\dagger \hat{Y}_{BS} \right> =
\left< \hat{b}_1^\dagger \hat{b}_1 \right> =
 t_1^2 \left| \alpha \right| ^2 + r_2^2 { \sinh}^2(s).
\label{eqn-bs-photon-number}
\end{eqnarray}

This results in an optimum {\lq\lq}phase resolution{\rq\rq} in the
output beam of

\begin{eqnarray}
S_{BS} &=& \left[
            {      \left< \hat{Y}_{BS}^\dagger \hat{Y}_{BS} \right>
                   \over
                   \left< \Delta \hat{X}_{BS}^2 \right>
            }
           \right] ^{1/2}
=          \left[
            {  t_1^2 \left| \alpha \right| ^2 + r_2^2 { \sinh}^2(s)
                   \over
               1 - r_2^2 \left[ 1 - { \exp}(-2s) \right]
            } 
           \right] ^{1/2}
.
\label{eqn-bs-optimal-signal-noise}
\end{eqnarray}

This is the best possible phase resolution that can be obtained in the
output for a coherent state in one input, and a squeezed vacuum (with
squeezing parameter $s$) in the other. It depends on both the reflectivity
$r_2$ and transmissivity $t_1$ of the beam-splitter, which control the
mixing proportions of the squeezed and coherent fields. Note that the
corresponding expression that can be derived for the other output mode
($\hat{b}_2$) is equivalent to this, the only difference being in the
optimum choice of phases.

An interferometer can also be used to give the squeezed vacuum some
coherent amplitude. Figure \ref{Fig4}(b) shows the layout schematically. 
The equations relating the input mode operators $\hat{a}_1$ and $\hat{a}_2$
to the output mode operators $\hat{c}_1$ and $\hat{c}_2$
are\cite{Caves-1981}

\begin{eqnarray}
\hat{c}_1 &=& { \exp}(i\Phi) \left[ 
    -i { \exp}(-i\psi) \hat{a}_1 { \sin}(\phi/2) + \hat{a}_2 { \cos}(\phi/2)
  \right] , \nonumber \\
\hat{c}_2 &=& { \exp}(i\Phi) \left[ 
    \hat{a}_1 { \cos}(\phi/2) - i { \exp}(-i\psi) \hat{a}_2 { \sin}(\phi/2) 
  \right]. 
\label{def-in-coupling}
\end{eqnarray}

For the same input fields that were used for the beam-splitter example
above, the output uncertainty quadrature $\hat{X}_{IN} = \hat{c}_1 +
\hat{c}_1^\dagger$ has a variance of

\begin{eqnarray}
\left< \Delta \hat{X}_{IN}^2 \right> = 
1 - \left[ 1 - { \exp}(-2s) \right] { \cos}^2(\phi/2). 
\label{def-in-variance}
\end{eqnarray}

The distance quadrature $\hat{Y}_{IN} = \hat{c}_1$, so

\begin{eqnarray}
\left< \Delta \hat{X}_{IN}^2 \right> = 
\left< \hat{c}_1^\dagger \hat{c}_1 \right> 
= \left| \alpha \right| 2 { \sin}^2(\phi/2) 
+ { \sinh}^2(s) { \cos}^2(\phi/2).
\label{eqn-in-photon-number}
\end{eqnarray}

The phase resolution in output mode $\hat{c}_1$ of the
interferometer is

\begin{eqnarray}
S_{IN}^2 &=&
   \left[
              {
                   \left< \Delta \hat{Y}_{IN}^2 \right>
                   \over
                   \left< \Delta \hat{X}_{IN}^2 \right>
              } 
   \right] ^{1/2}
\\
&=&  \left[
     {
            \left| \alpha \right| ^2 { \sin}^2(\phi/2) 
            + { \sinh}^2(s) { \cos}^2(\phi/2)
                   \over
            1 - \left[ 1 - { \exp}(-2s) \right] { \cos}^2(\phi/2)
     } 
   \right] ^{1/2}
.
\label{eqn-in-optimum}
\end{eqnarray}

This is the best possible phase resolution that can be obtained for a
coherent state in one input, and a squeezed vacuum (with squeezing
parameter $s$) entering the other. It depends on the relative phase shift
$\phi$ between the two arms of the interferometer. The relative phase
controls the proportions of the mixing between the squeezed and coherent
fields.  

\end{section}
\begin{section}{Squeezing efficiency}

To work out the overall energy efficiency of producing a light field with a
better phase resolution by mixing squeezed light with a coherent field, it is
necessary to consider how the two fields are created and mixed. A simple
theoretical scheme to create squeezed coherent light is shown in Fig. 
\ref{Fig5}. An initial pump coherent field (1) is passed through a parametric
oscillator to generate the squeezed vacuum (2). If the initial coherent field
has $N=\left| \alpha \right| ^2$ photons, the squeezed state contains $N_{sq} =
(N/2)^{1/2}$ photons. For $N$ large, the unconverted part of the coherent field
exiting the oscillator is largely unaffected, and this is then down-converted
(3) into $2 N \lambda$ signal frequency photons, where $\lambda$ is an
efficiency factor. This down conversion could be achieved using a parametric
amplifier.  The two fields at the sub-harmonic frequency are now mixed by
either the beam-splitter or the interferometer at (4). The inputs to this mixer
are a coherent state with $2 N \lambda$ photons, and a squeezed vacuum
containing $N_{sq}$ photons.  

\begin{figure}
\includegraphics[width=80mm]{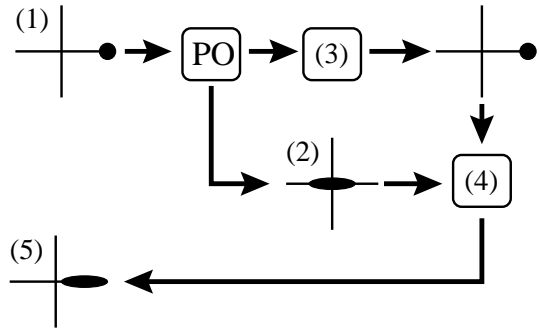}
\caption{
\label{Fig5}
Scheme to create squeezed light with a coherent amplitude from 
an initial coherent input (1). The input is fed into a 
parametric oscillator, producing a squeezed vacuum (2). The 
unconverted part of the coherent input is then down converted 
using another parametric system (3). The two signal frequency 
fields are then mixed at (4) by either a beam-splitter or an 
interferometer. The resulting output is a squeezed state with 
coherent amplitude (5).
}
\end{figure}

The number of photons in an ideal squeezed vacuum is $N_{sq} = \sinh^2(s)$,
so in the limit of large squeezing $\left| s \right| \gg 0$, and with
$N^{1/2} \gg 1$, we have $s \approx {1 \over 4} \ln (N/2)$; which is now
substituted into the equations for the phase resolution. For the ideal case
of a lossless beam-splitter with $t_1^2 = 1 - r_2^2$, the phase resolution 
from eqn (\ref{eqn-bs-optimal-signal-noise}) becomes 

\begin{eqnarray}
S_{BS} = 
2 N 
  \left[
    {
      \lambda \left( 1 - r_2^2 \right) + r_2^2 N^{-1/2} / \sqrt{8}
    \over
    1 - r_2^2 + r_2^2 N^{-1/2} / \sqrt{8}
    }
  \right] ^{1/2}
.
\label{eqn-bs-exact-ratio}
\end{eqnarray}

For large $N$ and small $r_2$ this can be approximated to

\begin{eqnarray}
S_{BS} 
&\approx& N^{1/2} \left( 2\lambda + 4 r_2^2 \right) ^{1/2}
\approx \left( 2 N \lambda \right) ^{1/2},
\label{eqn-bs-ratio} 
\end{eqnarray}

since $N^{-1/2} \ll 1$. The scaling of $S_{BS}$ remains the same, although
for $r_2 \approx 1$ the coefficient changes to become $S_{BS} \approx
(2N)^{1/2}$ since this is the limit corresponding to giving the squeezed
vacuum no coherent amplitude at all. A similar analysis with the same
approximations can be done for the interferometer. Using equation
(\ref{eqn-in-optimum}), the result is

\begin{eqnarray}
S_{IN} = 
2 N 
  \left[
    {
      \sqrt{8} \lambda { \tan}^2(\phi/2) + N^{-1/2} 
    \over
      \sqrt{8} { \tan}^2(\phi/2) + N^{-1/2} 
    }
  \right] ^{1/2}
.
\label{eqn-in-exact-ratio}
\end{eqnarray}

Just as for the beam splitter, for $\left| \phi - \pi/2 \right| \ll
1$ the first terms in the numerator and denominator dominate, so
that 

\begin{eqnarray}
S_{IN} \approx \left( 2 N \lambda \right) ^{1/2}
\label{eqn-in-ratio}
\end{eqnarray}

and for $\phi \ll 1/N$ the second terms dominate to give $S_{IN} \approx
(2N)^{1/2}$. However, as for the beam-splitter with $r_2 \approx 1$, the $\phi
\ll 1/N$ region corresponds to a field with negligible coherent amplitude, so
again no improvement in the scaling is achieved.
  
\begin{figure}
\includegraphics[width=80mm]{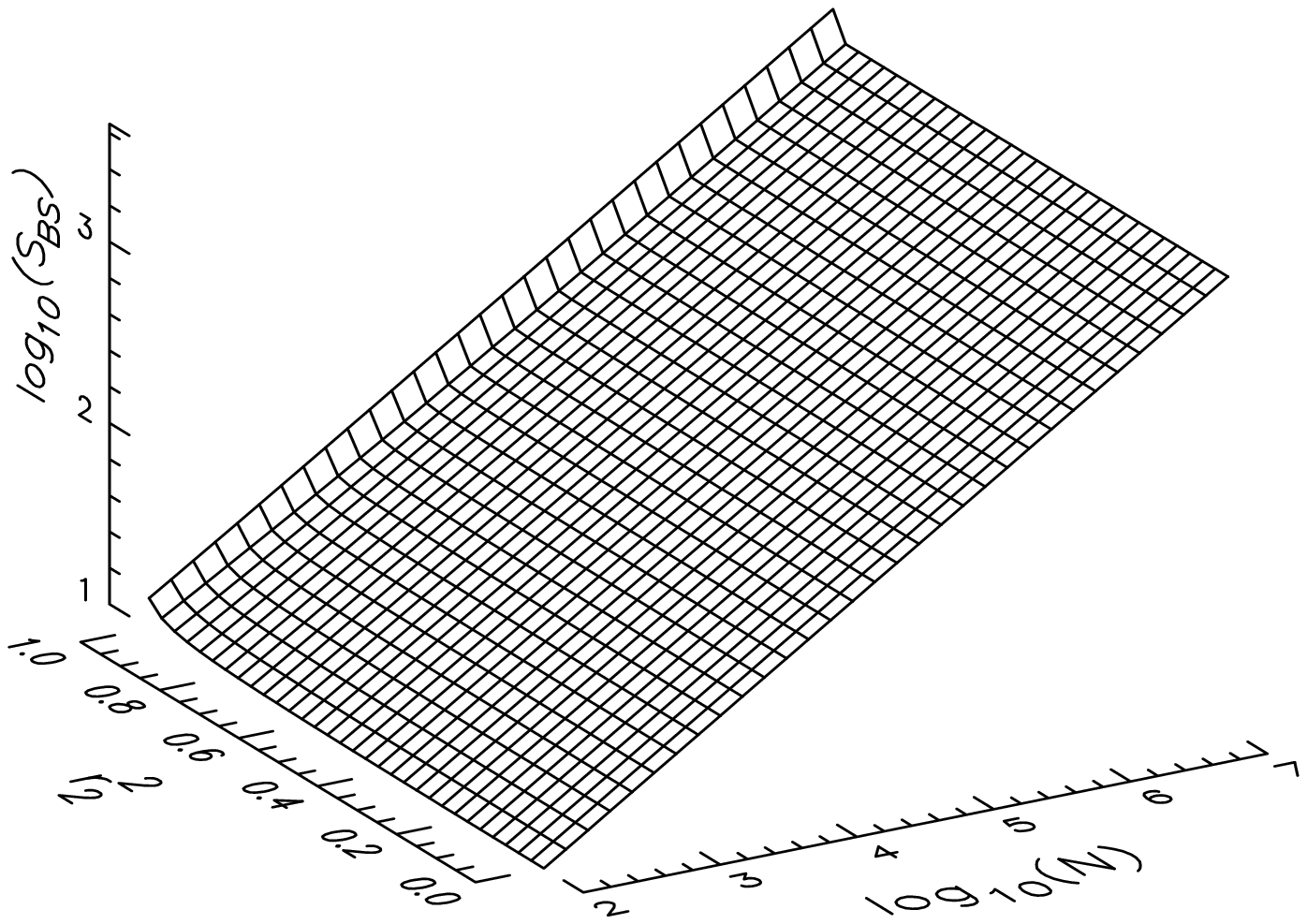}
\includegraphics[width=80mm]{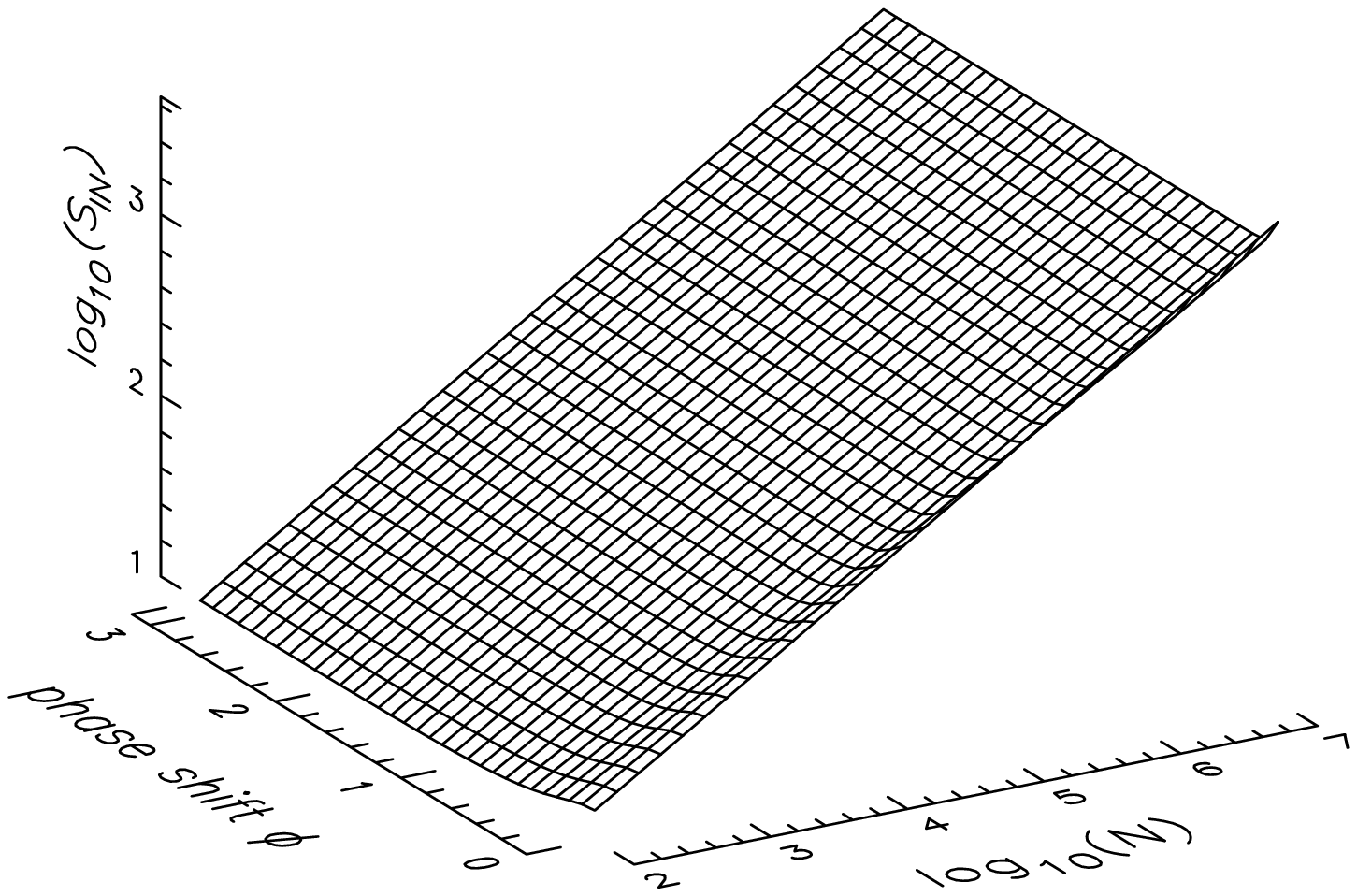}
\caption{
\label{Fig6}
Graphs of the phase "signal to noise" ratio for (a) a beam 
splitter and (b) an interferometer, as a function of input 
photon number and (a) reflectivity or (b) phase difference. 
The coherent squeezed light is generated using the scheme 
shown in the previous figure, and the coherent down 
conversion efficiency is $50\%$ ($\lambda=0.50$).
}
\end{figure}

The results are shown on Fig. \ref{Fig6}, which shows the phase resolution
defined in eqn (\ref{def-signal-to-noise}) for a squeezed vacuum with
$N^{1/2}$ photons mixed with a coherent state with $N$ photons.  Figure
\ref{Fig6}(a) shows the results from equation
(\ref{eqn-bs-optimal-signal-noise}) for a lossless beam-splitter -- the
$r_2$ axis goes from $r_2=0$ (passing only the coherent input), to $r_2=1$
(passing only the squeezed input). Figure \ref{Fig6}(b) shows the results
for the interferometer from equation (\ref{eqn-in-optimum}). The phase axis
$\phi$ varies from $\phi=0$, when only the squeezed input is passed into
the output mode $\hat{b}_1$; through to $\phi=\pi$, when only the coherent
input is passed.
  
\end{section}
\begin{section}{Conclusion}

For both the beam-splitter and the interferometer the scaling of the phase
resolution is identical, regardless of the relative mixing proportions
($r_2$ or $\phi$). In these terms, no benefit is obtained by mixing with a
coherent signal. The reason is easy to see, since letting part of the
coherent field into the output also lets through an equal proportion of its
vacuum fluctuations. This means the accuracy with which the phase of the
output of the system can be measured scales in the same way as the accuracy
with which the input can be measured. This result confirms the conjectures
of Kinsler, Fernee, and Drummond \cite{Kinsler-FD-1993} suggesting the
existence of a bound on {\lq\lq}phase information\rq\rq rates in systems
with a phase-invariant Hamiltonian. This is one step beyond a consideration
of a constant phase in a system, since it involved non equilibrium
processes.

Xaio et al \cite{Xiao-WK-1987} constructed a Mach-Zehnder
interferometer using squeezed light at the vacuum port. They reported an
improved measurement precision for the in-phase quadrature of the squeezed
light, but did not consider any type of phase measurement.  However, the
efficiency limit discussed here can just as easily be applied to schemes
involving quadrature measurements or fringe visibility.  For example, a
gravity wave interferometer would be set up to the measure the small
changes in length of the two arms caused by a passing gravity wave. In this
case, the relative phases of the light from the two arms would not be not
directly measured -- but the movement of the interference fringes would be.

Caves \cite{Caves-1981} has also considered the use of squeezed light in
interferometers. He found that partially squeezing the field going into the
vacuum port improved the sensitivity of the interferometer by reducing the
vacuum fluctuations added to the measurement. However, the large number of
photons in a highly squeezed input causes increased uncertainty from the
radiation pressure as individual photons reflect off the mirrors.  As a
result, the trade-off between reduced quantum noise and increased radiation
pressure means that the optimum sensitivity is obtained for a medium value
of the squeezing, at a lower input power. Caves's result did not beat the
standard quantum limit for the noise in the interferometer, and so the
benefit from his squeezed light scheme was that it operated at a lower
power. Above I showed that if the total available power is limited, then
the maximum amount of squeezing possible is also limited, whereas Caves
assumed that any desired amount of squeezing is available at the vacuum
input port.  Adding the squeezing efficiency constraint to the Caves
interferometer puts an even lower limit on the power reduction
possible with such a squeezed light scheme.

Subsequent schemes have shown that the standard quantum limit can be beaten
by either careful choice of the phase of the squeezed light
\cite{Unruh-1983,Bondurant-S-1984,Jaekel-R-1990,Luis-S-1992}, or by putting
a non linear Kerr medium in the arms of the interferometer
\cite{Bondurant-1986}. Taking into account the efficiency consideration
would not interfere with the lower noise in these schemes, but it does put
a limit on the squeezing available for a given maximum power. This will in
turn limit by how much the standard quantum limit can be beaten.  For
example, Pace et al \cite{Pace-CW-1993} constructed a model that included
mirror noise by describing the mirrors as harmonic oscillators coupled to
the light field. They found that by optimising the phase of the squeezed
light the standard quantum limit could be beaten. The optimum intensity
sensitivity was roughly proportional to $N_{sq}^{-1}$, the reciprocal of
the number of photons in the squeezed vacuum. However, as the squeezing is
increased, the detector sensitivity was found to be limited by the noise
caused by the mirrors so that beyond a certain point more squeezing had no
effect. This type of result had also been obtained by Jaekel and Reynaud
\cite{Jaekel-R-1990}, and Luis and Sanchez-Soto \cite{Luis-S-1992}.

If the mirror noise due to the radiation pressure is arbitrarily ignored,
and the optimum input power is used, Pace et al's \cite{Pace-CW-1993}
results show that the minimum detectable gravitational wave amplitude
$h_{min}$ is proportional to $e^{-s}$. If $P_{opt}$ is the optimum power,
which is the same as that for a non squeezed light scheme, and $s$ is the
squeezing parameter of the squeezed vacuum; then given the efficiency
constraint of $N^{-1/2}$ for the squeezed vacuum generated by the
parametric oscillator, this means than $h_{min}$ is proportional to
$P_{opt}^{1/4}$. However, it is still possible to use squeezing to
compensate for a lower, and no longer optimum, power. This is because the
decrease in sensitivity due to the characteristics of the light does not
matter if the measurements are still limited by the background noise caused
by the mirrors.

In summary, these results suggest that when using squeezed light generated
with passive time-independent parametric devices, the chief advantage is
not one of overall efficiency or improved sensitivity. Instead, the
advantage is one of lower power levels. Given a fixed incident coherent
pump power, the limits to the phase resolution available from passive
sources of squeezed radiation ensure that we cannot get improved
information capacity relative to the pump itself, nor is there any
improvement in phase resolution at a given pump photon number.  

\end{section}

\begin{section}*{Acknowledgments}
The author would like to thank P.D. Drummond and the Department
of Physics at the University of Queensland.  The paper has also gone
through revisions while I worked at the University of Sheffield and 
the University of Leeds, and an earlier version was at one stage submitted to 
Optics Communications but never published there.
\end{section}

\newpage 


\end{document}